\documentclass{emulateapj-rtx4}

\pdfoutput=1

\usepackage{lscape}
\usepackage{subfigure}
\usepackage{txfonts}
\usepackage{hyperref}
\usepackage{epsfig}
\usepackage{natbib}

\renewcommand{\anchor}[2]{\href{#1}{#2}}

\newcommand{\es}  {{erg s$^{-1}$}}

\defcitealias{liu01-11}{L11}
\defcitealias{hwang04-08}{HL08}

\newcommand{\HST}{{\em Hubble Space Telescope}}
\newcommand{\HSTt}{{\em HST}}
\newcommand{\Chandra}{{\em Chandra}}
\newcommand{\Chandraf}{{\em Chandra X-Ray Observatory}}

\newcommand{\ACIS}   {{ACIS}}
\newcommand{\CIAO}   {{\em CIAO}}

\newcommand{\CALDB} {{\em CALDB}}

\slugcomment{}
\shorttitle{Stacking Star Clusters in M51: Searching for Faint X-Ray Binaries}
\shortauthors{Vulic et. al.}

\begin{document}

\submitted{Accepted by: $\textit{The Astrophysical Journal}$}
\title{Stacking Star Clusters in M51: Searching for Faint X-Ray Binaries}

\author{N. Vulic\altaffilmark{1}\altaffilmark{$\bigstar$}, P. Barmby\altaffilmark{1}, and S. C. Gallagher\altaffilmark{1}}

\altaffiltext{1}{Department of Physics \& Astronomy, Western University, London, ON, N6A 3K7, Canada}
\altaffiltext{$\bigstar$}{$\texttt{nvulic@astro.uwo.ca}$}

\begin{abstract}
The population of low-luminosity ($<$ 10$^{35}$ \es) X-Ray Binaries (XRBs) has been investigated in our Galaxy and M31 but not further.
To address this problem, we have used data from the \Chandraf\ and the \HST\ to investigate the faint population of XRBs in the grand-design spiral galaxy M51. A matching analysis found 25 star clusters coincident with 20 X-ray point sources within 1.5$\arcsec$ ($60$ pc). From X-ray and optical color-color plots we determine that this population is dominated by high-mass XRBs. A stacking analysis of the X-ray data at the positions of optically-identified star clusters was completed to probe low-luminosity X-ray sources. No cluster type had a significant detection in any X-ray energy band. An average globular cluster had the largest upper limit, $9.23 \times 10^{34}$ \es, in the full-band ($0.3 - 8$ keV) while on average the complete sample of clusters had the lowest upper limit, $6.46 \times 10^{33}$ \es~in the hard-band ($2 - 8$ keV). We determined average luminosities of the young and old star cluster populations and compared the results to those from the Milky Way. We conclude that deeper X-ray data is required to identify faint sources with a stacking analysis.
\end{abstract}

\section{Introduction} \label{sec:intro}

An X-ray binary (XRB) consists of a compact object such as a neutron star (NS) or black hole (BH), that accretes matter from a donor star, which is generally a main sequence (MS) or red giant star.
Bright XRBs have average luminosities of $L_{x} \sim 10^{36-38}$ \es\ \citep{heinke02-09}. XRBs are divided into different categories based on the mass of the donor star: low-mass (LMXB) or high-mass (HMXB). There are other divisions made, which depend on system properties, that include X-ray bursters, X-ray pulsars, millisecond pulsars, and microquasars.

In LMXB systems, the donor stars are usually spectral type G-M with masses $\lesssim$ 1.5 M$_{\sun}$ \citep{hynes10-10}. As a result of both members being low-mass, LMXBs typically have periods of minutes to days \citep{charles08-03} and transfer mass predominantly via Roche lobe overflow. 
Due to the high stellar densities in globular clusters (GCs), LMXBs form via two and three-body encounters at a higher rate (per unit mass) than in the field \citep{katz02-75, clark08-75, fabian08-75, pooley07-03}. Various studies \citep{sarazin10-03, jordan09-04, kim08-06, sivakoff05-07, kundu06-07, humphrey12-08, kim09-09} found that $\sim 20\%-70\%$ of extragalactic LMXBs are located in GCs. GCs that are compact, massive, bright, and red (metal-rich) show a preference for hosting LMXBs \citep{sivakoff05-07, paolillo08-11}. Specifically, these studies indicate that red GCs are $\sim$ 3 times more likely to host LMXBs than blue GCs are.
Several explanations have been put forth to describe the metallicity effect. \citet{ivanova01-06} posited a process involving magnetic braking in main-sequence stars that would result in the suppression of LMXB formation in metal-poor GCs.
\citet{maccarone05-04} have suggested that irradiation-induced stellar winds are stronger in low-metallicity stars because emission line cooling is not efficient, speeding up LMXB evolution. This would result in shorter LMXB lifetimes and therefore a decrease in the number of LMXBs observed in metal-poor GCs.

In our Galaxy, \Chandra~has identified many low-luminosity ($L_{x}=10^{30-34}$ \es; \citealt{heinke10-05}) X-ray sources within GCs \citep{pooley07-03} such as millisecond radio pulsars and cataclysmic variables (CVs). Millisecond radio pulsars ($L_{x} \sim (1-4)\times10^{30}$ \es) are believed to be the final stage of NS-LMXB evolution whereby the NS stops accreting appreciable amounts of material and turns on as a radio pulsar. Studies of millisecond radio pulsars in GCs have confirmed the dynamical and metallicity effects on their formation in both the radio \citep{hui05-10} and gamma-ray \citep{abdo12-10, hui01-11} regimes.
With 22 discovered millisecond radio pulsars in GC 47 Tuc \citep{lorimer-03} and possibly twice as many undetected \citep{heinke10-05}, GCs are a hotspot for the study of LMXB evolution.

By contrast, HMXBs have donor stars of spectral type O or B with orbital periods of days, months, or possibly years \citep{lewin04-06}. HMXBs transfer mass via Roche lobe overflow and/or by Bondi-Hoyle (wind) accretion \citep{bondi-44, iben09-95}. Unlike LMXBs, HMXBs are not found to form preferentially in star clusters. Instead, there is an association between HMXBs and star-forming regions, where \citet{shtykovskiy05-07} found a low-significance ($\lesssim 2\sigma$) constraint on the (wider) distribution of HMXBs relative to bright HII regions. A study of three starburst galaxies by \citet{kaaret02-04} suggest that bright HMXBs are located near young clusters (YCs) of stars.
As a result of binary-binary or three-body interactions, coupled with the weak gravitational attraction to the cluster core, stars and compact objects can be easily ejected from YCs within a few Myr \citep{portegies-zwart08-99, kaaret02-04}. Different groups \citep{kalogera08-06, sepinsky03-05} have used the StarTrack population synthesis code \citep{belczynski01-08} in combination with a prescription for young cluster potentials to show that $\simeq$ 70\% of binaries (born within young cluster potentials) are ejected anywhere from $30-100$ pc as a result of supernova (natal) kicks for cluster ages of 15 Myr or older. The fact that few bright ($\sim$ 10$^{36}$ \es) XRBs are discovered near clusters is in line with observations of nearby starburst galaxies where ejections of $\gtrsim$ 200 pc \citep{rangelov11-11, kaaret02-04} have been found.

The role of dynamics in the dense environments of clusters is crucial in the evolution of binaries. For instance, not only do ejections illustrate the interactions occurring within these systems but \citet{vanbeveren07-12} state that binaries may have a significant effect on the chemical evolution of GCs. This link outlines the importance of binary evolution in stellar systems. Detailed theoretical modelling of binary evolution in clusters is complex and therefore requires observations to provide much-needed constraints. 
The emergence of a connection between the study of binary compact objects and gravitational waves \citep{belczynski06-02} is another source of motivation for uncovering the evolutionary processes of binary systems.

Before the launch of the \Chandraf\ in 1999, the study of X-ray sources was limited to the Local Group due to the low angular resolution and/or sensitivity of previous missions.
Much progress has been made investigating XRB populations in the last decade as a result of $\emph{Chandra's}$ subarcsecond resolution. However, there are still many unanswered questions. Currently, mostly bright XRBs are observed in other galaxies; M31 is the only nearby (large) galaxy where \Chandra~can observe sources below $L_{x} \sim 10^{35}$ \es\ \citep{zhang09-11}.
In general, since most elliptical galaxies are devoid of gas and dust (and therefore lack star formation), only LMXBs are formed and so it is straightforward to identify them.
The younger stellar population of spiral galaxies allows the formation of HMXBs, which have been identified separately from the diffuse emission in these galaxies via bright optical counterparts. 
Conversely, LMXBs have been more difficult to identify in spiral galaxies (because of faint optical counterparts) and therefore comprise an important element of future research.

M51 has been studied numerous times in X-rays by different groups, with the most recent population analysis from \Chandra\ observations by \citet{terashima02-04}. They focused on the bright-end of the X-ray luminosity function (XLF) of M51, investigating ultraluminous XRBs, and implied that these were likely stellar mass BHs accreting near/above $L_{Edd}$. 
This work was later extended by \citet{terashima07-06} and \citet{yoshida10-10} to detect optical counterparts and variability respectively. \citet{kilgard08-05} further studied M51's XLF (minimum point source luminosity of 2.2 $\times$ 10$^{37}$ \es) in comparison to other nearby spiral galaxies and found a flatter slope indicative of a larger HMXB population.
They also produced X-ray color-color diagrams in order to classify point sources (there is controversy about the use of this classification due to the spectral state variability of XRBs). Using these results, coupled with a study of variability, they concluded that a minimum of $\sim$ 27\% of all sources detected in a sample of 11 galaxies are accreting XRBs.
\citet{kilgard-06} performed a multi-wavelength analysis of M51's X-ray sources by matching them to optical data from the \HST\ (\HSTt). A total of 11 sources were matched to O or B-type stars and are likely HMXBs. They found 9 sources in young star clusters that seem to be a mix of XRBs and SNRs. \citet{maddox06-07} was able to match X-ray sources with radio observations and later confirmed 6 of the 24 sources identified in H$\alpha$ by \citet{kilgard-06} as SNRs.
An absence of optical counterparts coincident with M51b (also known as NGC5195, the companion galaxy interacting with M51) X-ray sources suggests that the population should be composed of LMXBs.
Therefore, by using multi-wavelength data, approximately half of the X-ray sources were optically classified as HMXBs or SNRs, with the remainder obscured or faint XRBs.

Besides deeper observations, there is a way we can probe the faint-end of the XLF and investigate low-luminosity XRBs. This is possible with a procedure known as stacking. By utilizing the methods of \citet{brandt07-01, brandt09-01} and \citet{hornschemeier06-01} we will stack star cluster positions in an X-ray image of M51 to produce in effect a deeper exposure of the average star cluster. By producing a final stacked image of many star clusters we hope to build up a source signal above the background level that would be evidence of faint sources. This will allow us to probe the low-luminosity end of the XLF and determine, on average, if star clusters in M51 host XRBs.
\citet{chandar08-04} found a mean $V-I$ color of M51 GCs identical to the peak of metal-poor Galactic GCs, where [Fe/H]$_{MW-GCs}$ = $-1.61$ \citep{burgarella05-01}. Therefore, the largely blue metal-poor GC population of M51 does not allow us a contrast with red metal-rich GCs. However, this makes a comparison with GCs in elliptical galaxies straightforward since they have near identical metallicities to M51's GCs \citep{chandar08-04}. By separating our clusters into different categories based on color (age) we investigate both HMXB and LMXB populations. Moreover, this analysis will help us better understand the metallicity (estimated from optical colors) dependencies of XRBs in both YCs and GCs.
In particular, observations that reveal the luminosity ranges and color correlations are crucial in constraining compact object and binary models such as the StarTrack population synthesis code \citep{belczynski01-08}.
More specifically, what do the average X-ray luminosities of YCs and GCs tell us about XRB populations?

We adopt a distance to M51 of 8.4 $\pm$ 0.6 Mpc determined by the planetary nebula luminosity function \citep{feldmeier04-97}. This agrees almost exactly with the recent determination of 8.4 $\pm$ 0.7 Mpc using supernovae 2005cs and 2011dh \citep{vinko11-11}.

\section{Observations}
\subsection{X-ray Data}

\Chandra~has observed M51 three times since 2000: 2000 June 20, 2000 June 23, and 2003 August 7, for exposure times of 15.06 ks, 27.15 ks, and 48.61 ks respectively. The X-ray point source catalogue that we use for this analysis was taken from \citet[hereafter L11]{liu01-11} and corresponds to that associated with the 48.61 ks observation (ObsID 3932).
This observation was obtained with $\emph{Chandra's}$ Advanced CCD Imaging Spectrometer (\ACIS) instrument in (TIMED) VFAINT mode with the galaxy located on the back-illuminated S3 chip (proposal 04600406, P.I.: Yuichi Terashima). \citetalias{liu01-11} identified 109 X-ray point sources. Taken with the longest exposure time, these data give us the most complete catalogue of X-ray point sources in M51 to date. For reference, in the $0.3 - 8$ keV energy band the limiting luminosity in the X-ray catalogue was $L_{x} \sim 4.6\times10^{36}$ \es\ and the maximum was $L_{x} \sim 5.4\times10^{39}$ \es\ while the average of all point sources was $L_{x} \sim 2.4\times10^{38}$ \es.

\subsubsection{Data Reduction} \label{sec:xraydata}

The 48.61 ks data were reprocessed with the updated \Chandra~Interactive Analysis of Observations (\CIAO) tools package version 4.3 \citep{fruscione07-06} and the \Chandra~Calibration database (\CALDB) version 4.4.6 \citep{graessle07-06} to clean the image for optimal analysis of faint sources. Starting from the level-1 events file, $\texttt{acis\_run\textunderscore hotpix}$ was used to produce a bad pixel file that identified observation-specific bad pixels, hot pixels, bright bias pixels, and afterglow events. Since the observation was completed in VFAINT mode and we are studying sources with a small number of counts, $\texttt{acis\_detect\_afterglow}$ was used to eliminate afterglows with very few events. Next, we ran $\texttt{acis\_process\_events}$ to update the level-1 events file for grade and status information, charge transfer inefficiency (CTI), time-dependent gain, and pulse height. The energy-dependent subpixel event repositioning (EDSER) algorithm parameter (under the pixel adjustment option) was used to improve the astrometric accuracy of our images.

Each \CIAO\ tool has unique processing parameters that need to be adjusted based on the given dataset.
Specifically, for VFAINT data, $\texttt{acis\_process\_events}$ was set to use a 5 $\times$ 5 event island to identify cosmic-ray background events as opposed to the usual 3 $\times$ 3 block. After filtering using the standard (ASCA) grades ($0, 2-4, 6$), status bits ($0$), and the good time intervals, we produced images in various energy bands. These include fully processed soft, hard, and full-band images in the energy ranges of $0.3-2$ keV, $2-8$ keV, and $0.3-8$ keV respectively.
Energies above 8 keV are not used since the effective area of the \Chandra~High Resolution Mirror Assembly steeply decreases with energy while the background increases (e.g., \citealt{brandt12-01}).
For each image we computed an exposure map using the \CIAO\ tools $\texttt{asphist}$, $\texttt{mkinstmap}$, $\texttt{get\_sky\_limits}$, and $\texttt{mkexpmap}$, which are included in a provided script called $\texttt{fluximage}$. We used monoenergies of $\sim$ 0.93 keV, $\sim$ 4.7 keV, and $\sim$ 2.1 keV,
which were the average energies for the soft, hard, and full band images respectively. These energies were used to calculate the instrument map as opposed to using weighted spectrum files.
Using the \CIAO\ tool $\texttt{dmimgcalc}$, we divided each image [counts pixel$^{-1}$] by its corresponding exposure map [cm$^{2}$ s counts photon$^{-1}$] to obtain final fluxed images in units of photons cm$^{-2}$ s$^{-1}$ pixel$^{-1}$. Each image has dimensions of approximately 9.7$\arcmin$ $\times$ 9.7$\arcmin$. Using the distance to M51 of 8.4 $\pm$ 0.6 Mpc, we determine a linear scale of 40.7 pc arcsecond$^{-1}$. The plate scale of the \Chandra\ images is 0.5 arcsecond pixel$^{-1}$, which corresponds to $\sim$ 20 pc pixel$^{-1}$.

In order to perform our stacking analysis accurately, both optical and X-ray images were registered to the 2MASS point source catalogue \citep{skrutskie02-06} reference frame. A total of 99 reference stars were found within the field of view of the \Chandra~image from the 2MASS catalogue. Using the program $\emph{immatch}$ in the WCSTools package, we found 6 reference stars that matched within a tolerance of 0.5$\arcsec$ of X-ray source counterparts.
This resulted in applying astrometric shifts of $-$0.210$\arcsec$ in right ascension and $-$0.073$\arcsec$ in declination to the image and to the point source catalogue compiled by \citetalias{liu01-11}.

\subsection{Optical Data} \label{sec:opt}
The star cluster catalogue used in this analysis, taken from \citet[hereafter HL08]{hwang04-08}, was compiled from data taken by the
$\emph{HST's}$ Advanced Camera for Surveys (ACS) with the F435W, F555W, F814W, and F658N filters. The effective exposure times were 16320 s, 8160 s, 8160 s, and 15640 s respectively 
(proposal 10452, P.I.: Steven V. W. Beckwith). Hereafter, B, V, and I will be used to denote colors corresponding to B$_{F435W}$, V$_{F555W}$, and I$_{F814W}$ respectively. The plate scale of the \HSTt\ ACS images is 0.05 arcseconds pixel$^{-1}$, which corresponds to $\sim$ 2 pc pixel$^{-1}$. The image has dimensions of approximately 6.8$\arcmin$ $\times$ 10.5$\arcmin$. We use all 2224 star clusters that have been identified by \citetalias{hwang04-08} as Class 1, meaning they are nearly circular and have no prominent nearby neighbours.

As a result of the poor astrometric accuracy of the \HSTt\ guide star catalogue, significant shifts are introduced in \HSTt\ images. We follow the same procedure as described above for determining the offset in the X-ray image. From the 2MASS catalogue we found a total of 98 reference stars in the \HSTt\ image.
The \HSTt\ offset was adjusted by applying astrometric shifts of $-$0.116 $\pm$ 0.320$\arcsec$ in right ascension and $-$0.653 $\pm$ 0.441$\arcsec$ in declination to the star cluster catalogue compiled by \citetalias{hwang04-08}. These values were calculated by matching 21 reference stars within a tolerance of 0.9$\arcsec$ of image stars.

\section{Data Analysis}
\subsection{X-ray Properties of Matched Star Clusters} \label{sec:xrmatches}
\begin{figure}[!ht]
\plotone{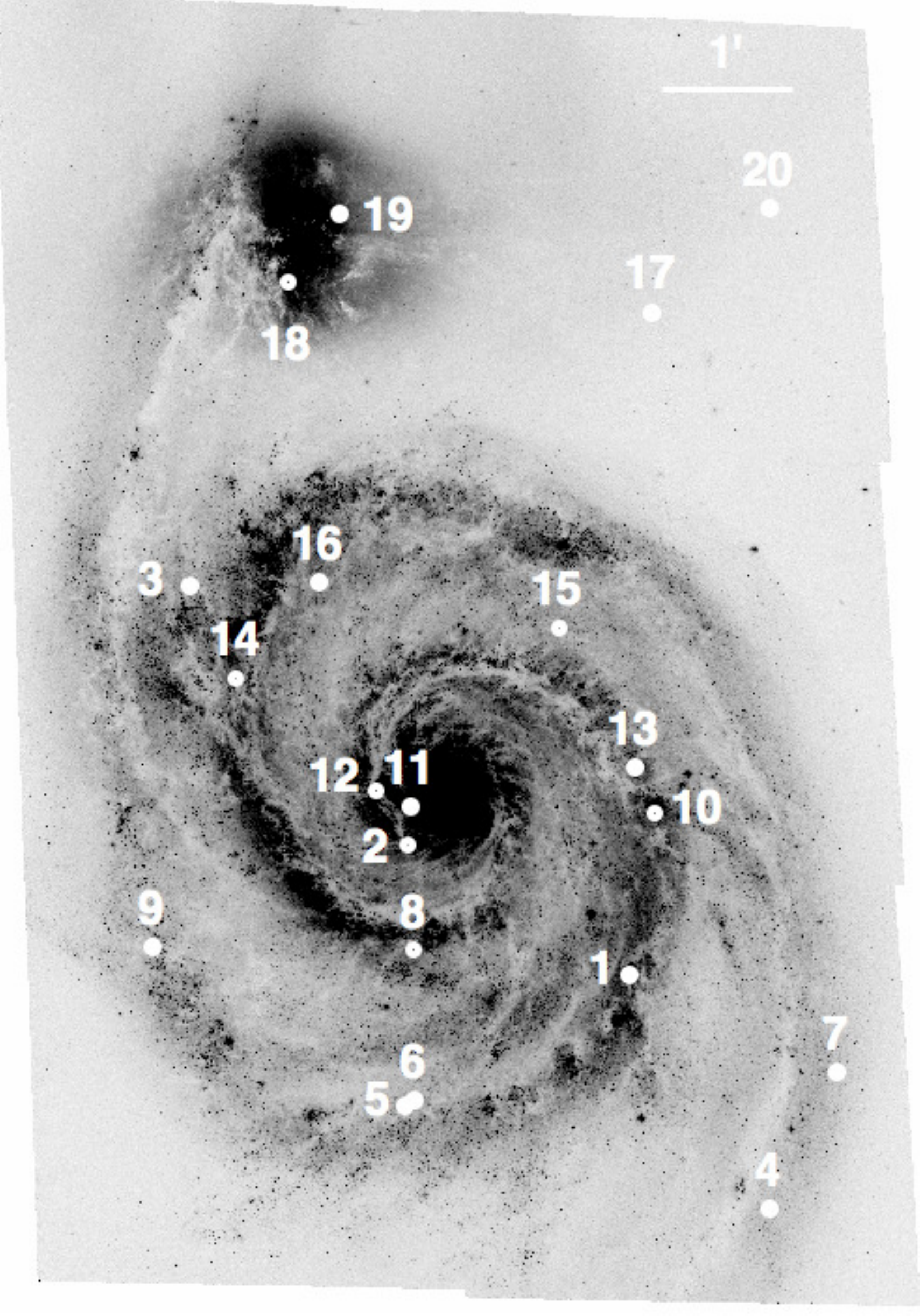}
\caption{\HST\ ACS image of M51 in the F435W filter showing star clusters matched to X-ray point sources using TOPCAT. Table \ref{tab:scxrmatches} has position and separation details including star cluster group information not shown here (see Figure \ref{fig:scgroups}).}
\label{fig:m51optical}
\end{figure}

\begin{figure}[!ht]
\begin{center}
\epsscale{0.35}
\begin{tabular}{ccc}
\plotone{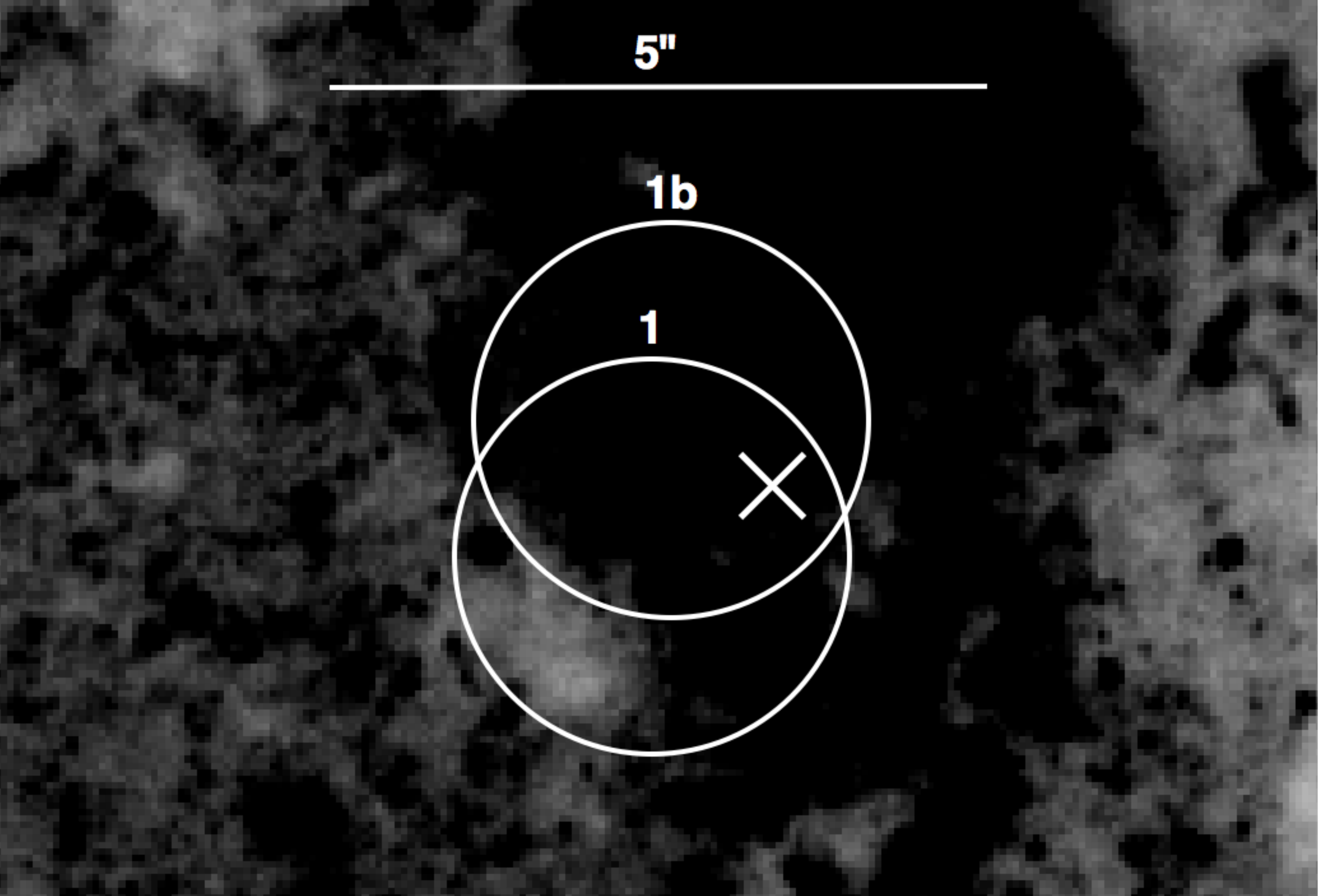}
\plotone{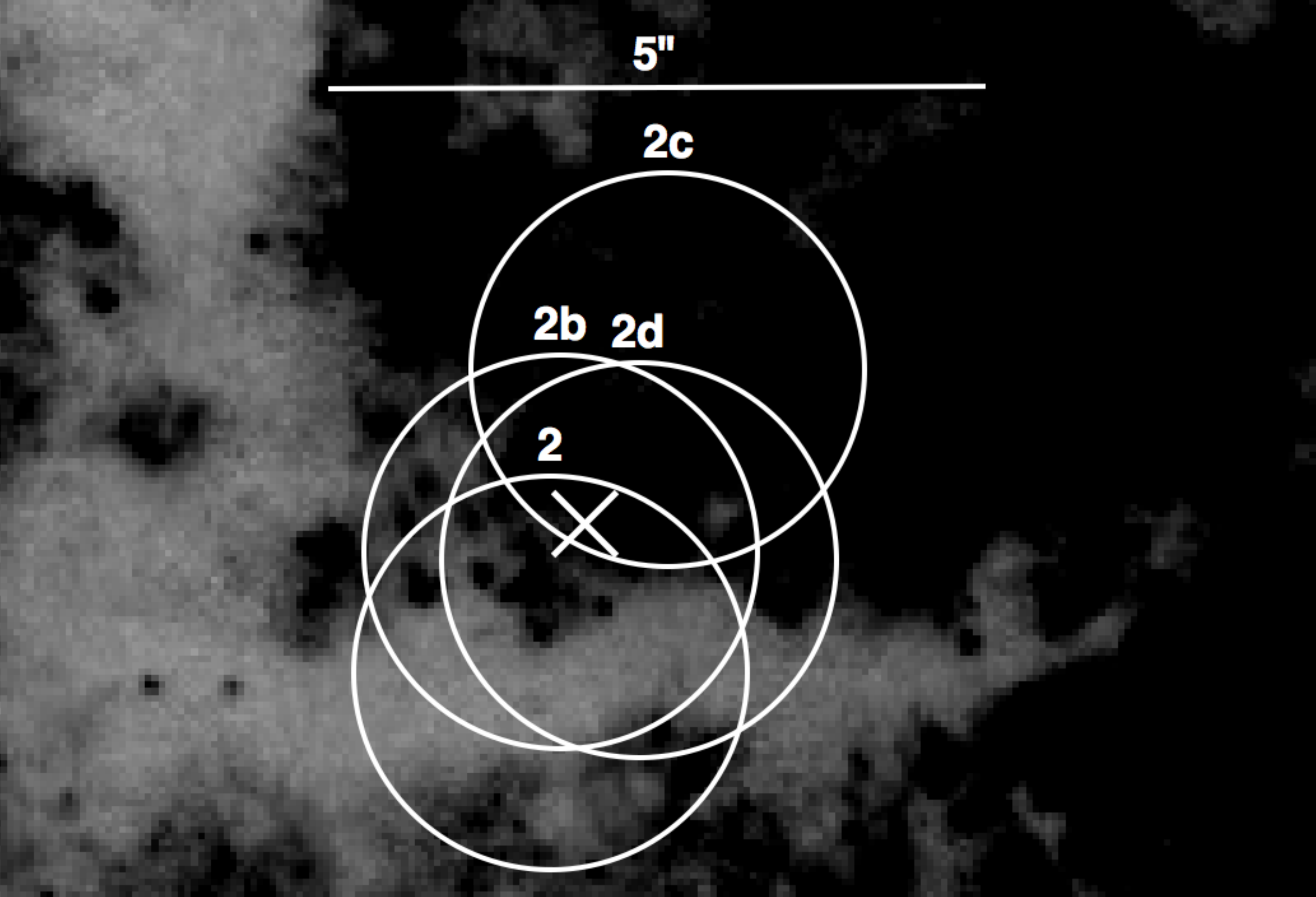}
\plotone{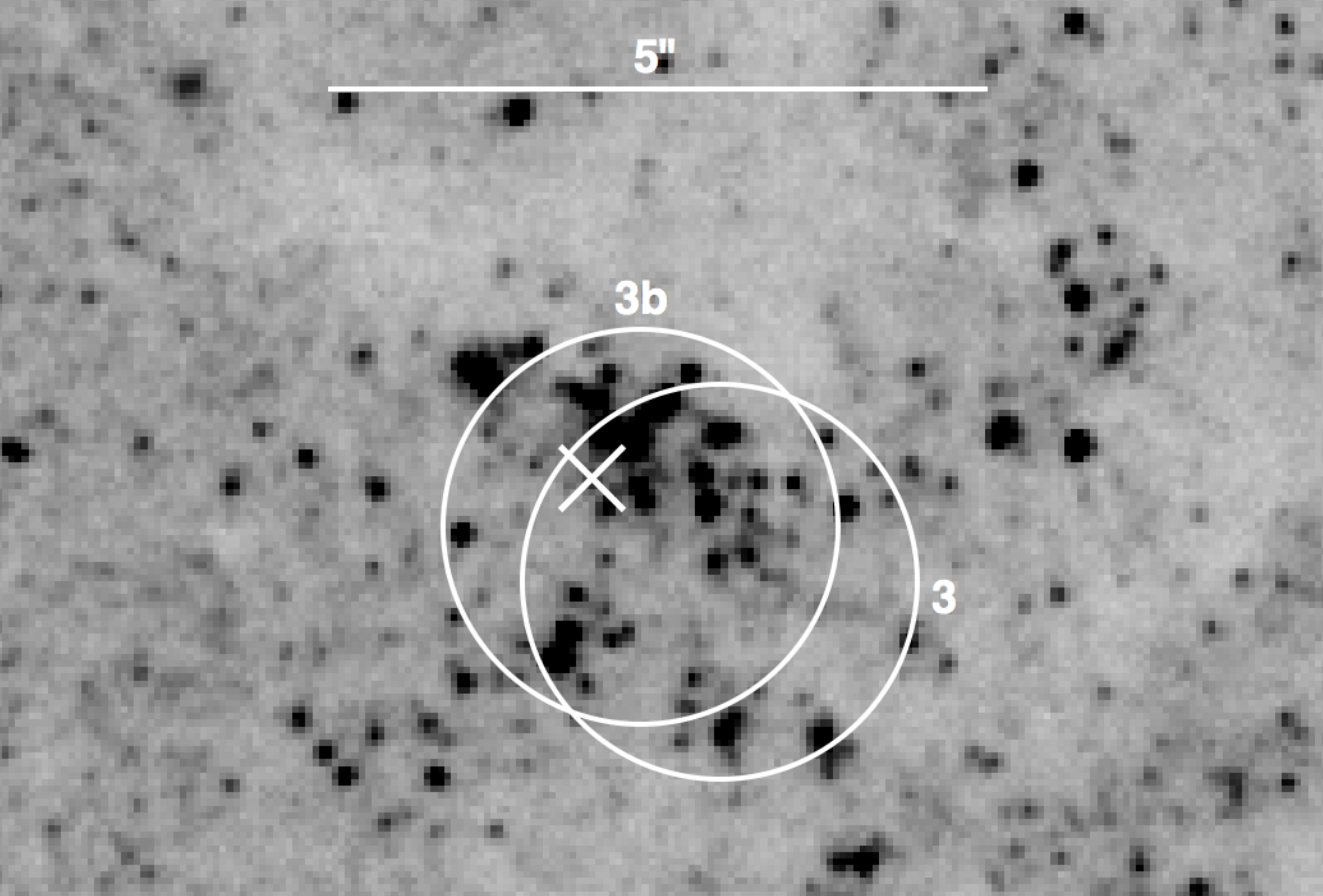}
\end{tabular}
\caption{Close-up of the star clusters from Figure \ref{fig:m51optical} that were identified as groups in Table \ref{tab:scxrmatches}. From left to right, cluster group IDs from Table \ref{tab:scxrmatches} are 1, 2, and 3 respectively. All panels have the same dimensions (10$\arcsec$ across by 6.7$\arcsec$ high) and all circles are 3$\arcsec$ in diameter. The X in each panel marks the location of the X-ray point source matched to each cluster within 1.5$\arcsec$.}
\label{fig:scgroups}
\end{center}
\end{figure}

\begin{figure}[!ht]
\plotone{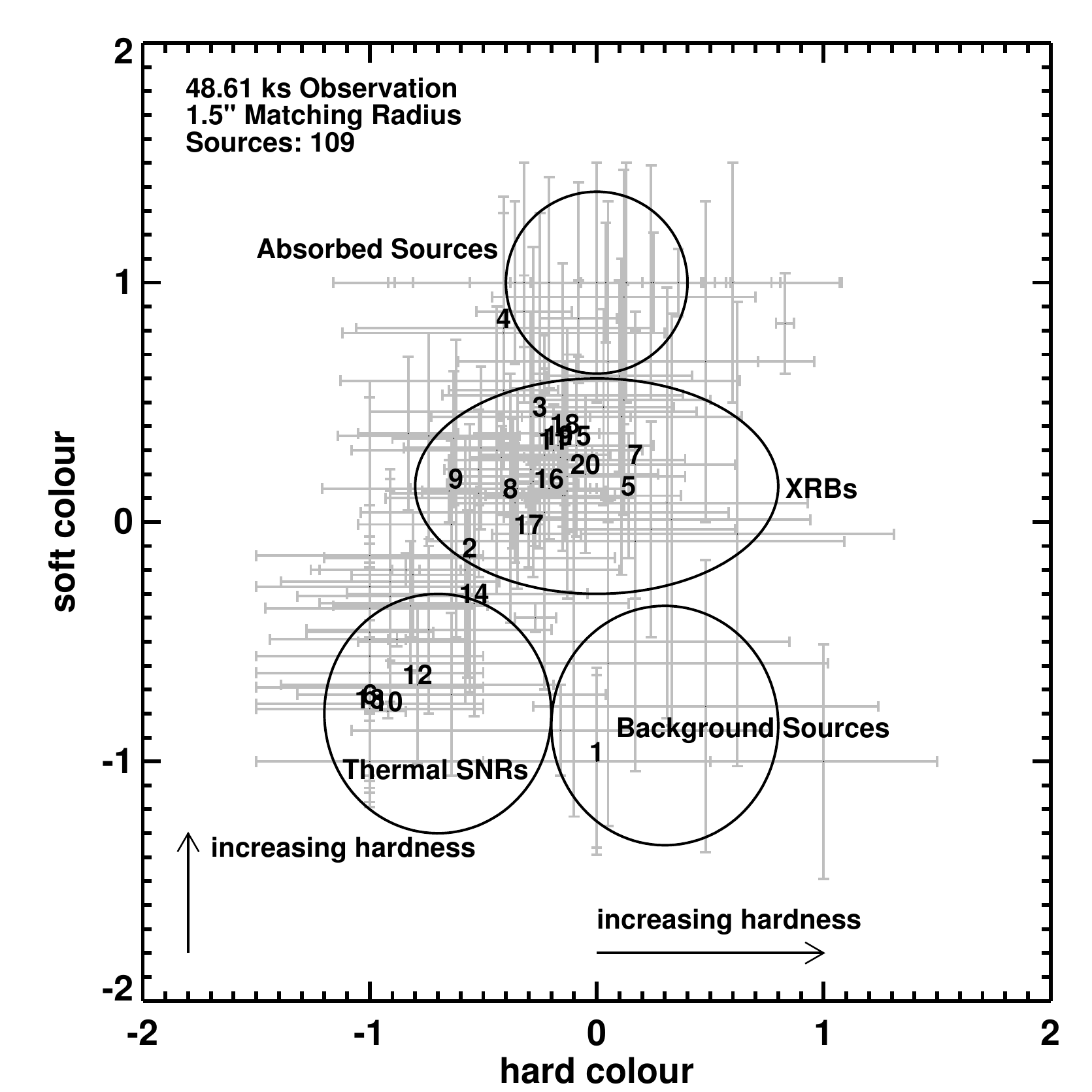}
\caption{X-ray color-color diagram of 109 X-ray point sources in M51 from \citetalias{liu01-11}. Of these point sources, 20 were matched within 1.5$\arcsec$ to 25 optically identified star clusters. The X-ray color classification scheme of \citet{kilgard08-05} (modified from \citealt{prestwich10-03}) is shown with the values in color space outlined in Table \ref{tab:colvals}.}
\label{fig:xraycc}
\end{figure}

In order to perform our stacking analysis, we first needed to determine if there were any direct matches of star cluster positions to X-ray point sources. We performed a matching analysis with TOPCAT \citep{taylor12-05} using the centres of each of the 2224 identified star clusters \citepalias{hwang04-08} with the 109
X-ray point sources identified by \citetalias{liu01-11}. We found 20 matches to 25 star cluster positions (multiple star clusters can be associated with one X-ray point source) within a 1.5$\arcsec$ radius. We used a slightly larger matching radius than previous studies (generally $\sim$ 1\arcsec) such as \citet{sivakoff06-08}. This value was used because we are observing mostly young clusters, where ejections are more likely, as opposed to GCs. The results are shown in Table \ref{tab:scxrmatches} along with an optical image of the matches in Figure \ref{fig:m51optical}. A close-up of the star cluster groups outlined in Table \ref{tab:scxrmatches} is shown in Figure \ref{fig:scgroups}. To assess the chance coincidence probability of our matches, we followed a method similar to that used by \citet{antoniou06-09} and \citet{zezas10-02}. By creating 25 source lists from our original X-ray catalogue with each source offset by $\pm$ 5$-$10$\arcsec$, we determined a false-match rate of $\sim 50\%$. This could be expected with a large matching radius and sample size. Using the 1$\arcsec$ matching radius our false-match rate is reduced to $\approx 20\%$. A color-color plot of the X-ray point sources is shown in Figure \ref{fig:xraycc}. Overlaid is the X-ray color classification scheme of \citet{kilgard08-05}, which was modified from \citet{prestwich10-03}. The values in color space are outlined in Table \ref{tab:colvals}.
XRBs have been combined into one group since delineating the population as low-mass or high-mass is difficult and controversial. Although this color classification is somewhat arbitrary (an XRB can end up in different regions of the diagram depending on its spectral state), it is a good approximation to the types of sources we are observing. In addition, the uncertainties associated with X-ray colors are large because of the low counts per source,
increasing the probability of possible misclassification.
Soft and hard colors (SC \& HC) are defined in equations \ref{sc} and \ref{hc} respectively based on the soft (S, $0.3-1$ keV), medium (M, $1-2$ keV), and hard (H, $2-8$ keV) band energies consistent with \citetalias{liu01-11}.
\begin{eqnarray}
SC & = & \frac{(M-S)}{(H+M+S)}
\label{sc}
\end{eqnarray}
\begin{eqnarray}
HC & = & \frac{(H-M)}{(H+M+S)}
\label{hc}
\end{eqnarray}

The majority ($\sim$ 65\%) of sources in Figure \ref{fig:xraycc} are classified as XRBs. Figure \ref{fig:m51optical} shows that most matches are located along the spiral arms of M51 and are likely HMXBs. The exceptions are matches $17-20$, which appear in M51b or are found at large radii from the nucleus of M51. Matches 18 and 19 may be bright LMXBs since \citet{kilgard-06} indicated a lack of optical counterparts in M51b. Match 19 is located 6\arcsec\ from the co-ordinates of SN1945A\footnote{Obtained from the NASA/IPAC Extragalactic Database (NED)} but would not be associated with emission from the remnant. If we assume that the ejecta from SN1945A has a similar velocity to that of SN1987A, the SNR would have a radius of only $\sim$ 0.5\arcsec\ (natal kick to the compact object is negligible). The remaining sources are mainly classified as SNRs (matches 6, 10, and $12-14$) and all reside within knots of the spiral arms, adding confidence to their classification.

\begin{deluxetable}{ll}
\tabletypesize{\footnotesize}
\tablecaption{X-ray Color-Color Classifications \label{tab:colvals}}
\tablecolumns{2}
\tablewidth{0pt}
\tablehead{\colhead{Classification} & \colhead{Range}}
\startdata
Supernova Remnant & HC $<$ $-$0.2, SC $<$ $-$0.3 \\
X-ray Binary & $-$0.8 $<$ HC $<$ 0.8, $-$0.3 $<$ SC $<$ 0.6 \\
Background Source & HC $>$ $-$0.2, SC $<$ $-$0.3 \\
Absorbed Source & SC $>$ 0.6 \\
Supersoft Source & M=0, H=0 \\
\enddata
\tablecomments{Classification scheme of \citet{kilgard08-05}.}
\end{deluxetable}

\subsection{Optical Properties of Matched Star Clusters} \label{sec:opt-match}
\begin{figure}
\plotone{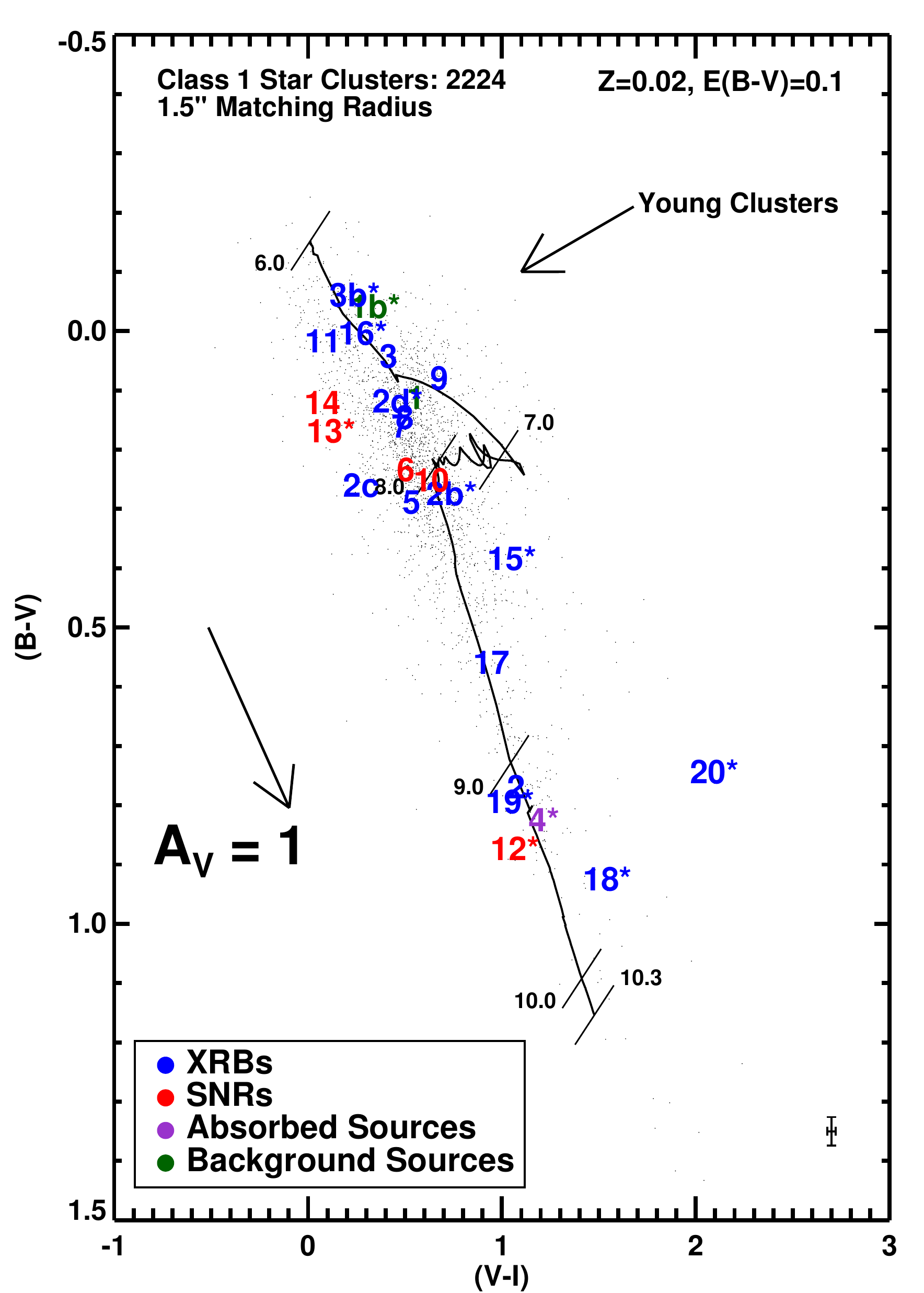}
\caption{Optical color-color diagram of the 2224 star clusters in M51 taken from the catalogue of \citetalias{hwang04-08}. A theoretical evolutionary track from the Simple Stellar Population (SSP) model of \citet{bruzual10-03} with $Z = 0.02$ is included with internal reddening of $E(B-V) = 0.1$ and foreground reddening $E(B-V) = 0.035$ applied. Labels from 6.0 to 10.3 represent log(age) in years. Numbers from $1-20$ are color-coded by X-ray type as indicated in the legend and correspond to the 25 matches to 20 X-ray point sources from \citetalias{liu01-11}. Numbers with an asterisk correspond to matches with separations $\leq 1\arcsec$ from Table \ref{tab:scxrmatches}. The foreground reddening vector is of length $A_{V} = 1$ mag. We have excluded clusters in the figure with $(B-V) > 1.5$ appearing well below the billion-year mark for illustration purposes. Matches 2, 12, 18, and 19 may all be heavily reddened (see Figure \ref{fig:m51optical}) and therefore we may expect them to move into the young region of color space. Match 20 may be an absorbed background source (AGN) because it is located far from the extent of the spiral arms (Figure \ref{fig:sccc}) and is quite red in $V-I$.}
\label{fig:sccc}
\end{figure}

\begin{figure}
\plotone{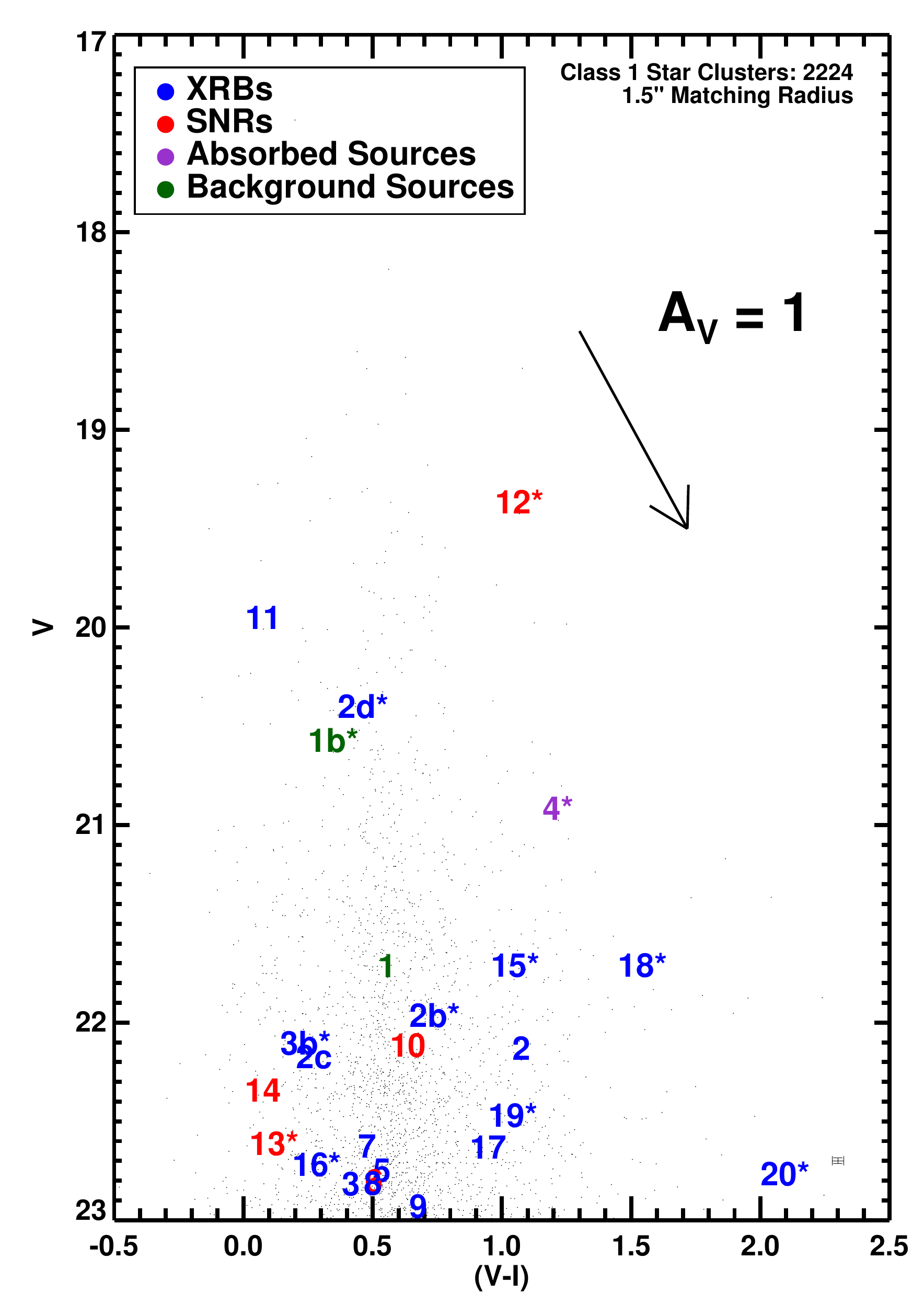}
\caption{Color-magnitude diagram of 2224 star clusters in M51 taken from the catalogue of \citetalias{hwang04-08}. Numbers from $1-20$ are color-coded by X-ray type as indicated in the legend and correspond to the 25 matches to 20 X-ray point sources from \citetalias{liu01-11}. The foreground reddening vector is of length $A_{V} = 1$ mag. Numbers with an asterisk correspond to matches with separations $\leq 1\arcsec$ from Table \ref{tab:scxrmatches}.}
\label{fig:cmd}
\end{figure}

The optical data for the 2224 star clusters in M51 \citepalias{hwang04-08} is used to make the color-color plot in Figure \ref{fig:sccc}. A theoretical evolutionary track from the Simple Stellar Population (SSP) of \citet{bruzual10-03} with $Z = 0.02$ is overlaid after applying internal reddening of $E(B-V) = 0.1$. The internal reddening was computed using $A_{V_{M51}} = 0.31$ mag from \citetalias{hwang04-08}.
Foreground reddening of $E(B-V) = 0.035$ from the dust maps of \citet{schlegel06-98} has also been applied to the SSP model.
The 25 labelled star clusters correspond to the 20 X-ray point sources they were matched to from Figure \ref{fig:xraycc}. Star cluster colors were calculated by \citetalias{hwang04-08} using a 6-pixel aperture in each band.
The SNRs indicated are located in clusters that are younger than 100 million years with the exception of match 12. This is likely a result of the increased reddening near the nucleus of M51 ($E(B-V) = 0.25 - 0.3$; \citealt{lamers02-02, kaleida08-10}) where match 12 is located, which would force it to move into the younger region of color space (refer to reddening vector in Figure \ref{fig:sccc}). The same may be true for matches 2, 18, and 19, which are all located in heavily obscured regions. Therefore, most XRBs are classified as HMXBs since they are associated with clusters younger than 100 Myr. Match 20 may be an absorbed background source (AGN) because it is located far from the extent of the spiral arms and has red $V-I$ colors in Figure \ref{fig:sccc}.
To determine whether our matches correspond to the brightest star clusters, we created the color-magnitude diagram show in Figure \ref{fig:cmd}. Though previous studies of XRBs in elliptical galaxies have found that they are preferentially found in luminous clusters \citep{sivakoff05-07, paolillo08-11}, our matches do not seem to favour this relation as they are not found within a particular region in color-magnitude space. However, the matches located in the most luminous clusters would be more likely to be associated with an X-ray point source (because luminous clusters are relatively rare), whereas clusters on the outskirts may be spurious matches. Many of the faint clusters were those mentioned to reside in regions of high extinction and could therefore be highly obscured as opposed to false-matches.

It is important to compare our matching results to previous studies to determine if any similarities exist among star cluster X-ray sources in different galaxies. \citet{sivakoff05-07} found 270 of 6488 ($\sim$ 4\%) GCs in 11 Virgo Cluster early-type galaxies with X-ray emission. This is in contrast to the 25 of 2224 ($\sim$ 1\%) of star clusters that show X-ray emission in M51. However, we used a matching radius that is 0.5$\arcsec$ larger than that used by \citet{sivakoff05-07}. Using this smaller radius, we find that only $\sim$ 0.5\% of star clusters are matched with X-ray point sources. This much lower value could be a consequence of the differing XRB populations of spiral and elliptical galaxies. While we have identified mainly HMXBs in M51's YCs, early type galaxies' XRB populations are comprised of LMXBs in GCs, allowing more time for them to form \citep{bogdan03-10}.
Combined with the fact that HMXBs are short-lived, we might expect a larger fraction of star clusters with X-ray emission in elliptical galaxies.

\section{X-Ray Stacking} \label{sec:stacking}

To further investigate the X-ray source population, specifically that of faint or quiescent XRBs, we performed a stacking analysis of the 2199 unmatched star clusters in M51. Following the methods of \citet{brandt07-01, brandt09-01} and \citet{hornschemeier06-01} we stacked 21 pixel by 21 pixel regions centred on star cluster positions in X-ray images of M51. This was completed using code written in the Interactive Data Language (IDL).
\subsection{Method \& Results} \label{sec:results}

\begin{figure*}
\plotone{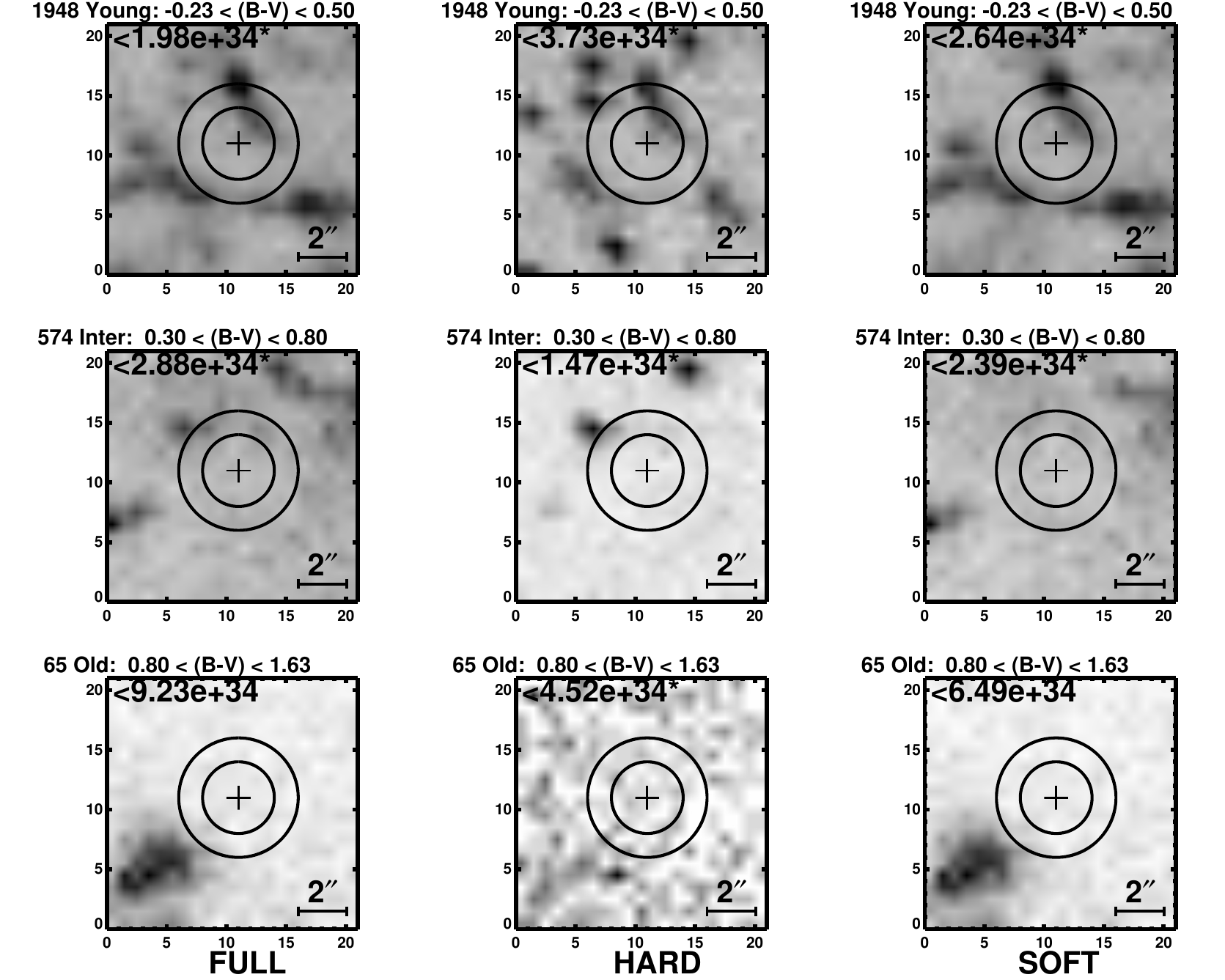}
\caption{{Stacked \Chandra~images of various star cluster types in M51. Above each image the cluster type and $B-V$ color range (from Table \ref{tab:scgroups}) is indicated along with the number of clusters stacked. Each column corresponds to images stacked in the full ($0.3-8$ keV), hard ($2-8$ keV), or soft ($0.3-2$ keV) energy bands.
The images are 21 pixels per side with each pixel 0.492$\arcsec$. North is up and east is to the left. The source aperture is 6 pixels in diameter while the background was calculated outside the larger aperture of diameter 10 pixels. Values in the top left corners of images indicate the 1.3$\sigma$ upper limit per cluster in the stack on the source luminosity in \es, where an asterisk denotes a negative luminosity in the source aperture (below the background level). The grayscale denotes the brightest pixel (darkest) relative to the dimmest pixel (lightest). All stacked images across all energy bands correspond to non-detections.}}
\label{fig:stack}
\end{figure*}

Before proceeding with stacking, all star cluster positions that were closer than 11 pixels to the edge of the X-ray image were excluded because a stacked image (21 by 21 pixels) requires this minimum number in each direction from the centre to be complete. This left a total of 2187 star clusters in the entire sample. We then subdivided star clusters into 5 different groups based on $B-V$ color as shown in Table \ref{tab:scgroups}. Each group was stacked in the full ($0.3-8$ keV), hard ($2-8$ keV), and soft ($0.3-2$ keV) energy bands. Our initial method consists of simply stacking all star cluster groups with no modifications to the star cluster regions in the X-ray image. The results are shown in Figure \ref{fig:stack}.

\begin{deluxetable}{lll}
\tabletypesize{\footnotesize}
\tablecaption{Stacked Star Cluster Groups \label{tab:scgroups}}
\tablecolumns{3}
\tablewidth{0pt}
\tablehead{\colhead{Group Type} & \colhead{Number} & \colhead{Optical color Range}}
\startdata
All Clusters & $2187$ & $-0.23 < (B-V) < 1.63$ \\
Very Young Clusters & $305$ & $-0.23 < (B-V) < 0.05$ \\
Young Clusters & $1948$ & $-0.23 < (B-V) < 0.25$ \\
Intermediate Clusters & $574$ & \phs$0.30 < (B-V) < 0.75$ \\
Old Clusters & $65$ & \phs$0.75 < (B-V) < 1.63$ \\
\enddata
\end{deluxetable}

The net counts in the source aperture (C$_{n}$) were determined through the relation $C_{s} - C_{b}\times A_{s}/A_{b}$, where C and A represent the counts and area respectively for the source (s) and background (b) regions. Our uncertainties were calculated using poisson statistics (method of \citealt{gehrels04-86}). We first computed the 3$\sigma$-clipped mean $\mu$ ($C_{b}/A_{b}$) in the background, where any pixels with counts greater than $\mu$ + 3$\sigma$ were excluded.
The upper and lower limits for the source and background region counts were calculated using equations (9) and (14) from \citet{gehrels04-86}.
We used a value of $S = 1.282$ (number of Gaussian $\sigma$) with confidence level parameters of $\beta = 0.01$ and  $\gamma = -4.0$.
The uncertainties in C$_{n}$ were determined by normalizing the background values to that of the source aperture.
Luminosities were computed from fluxes assuming a distance of 8.4 Mpc.
The uncertainties in luminosity were determined by taking a ratio of the luminosity in the source aperture to the counts in the source aperture and multiplying that ratio by the uncertainties in counts.

Our stacking yielded non-detections for all cluster types and energy bands. In most cases, except the soft and full-band old clusters, the background-subtracted luminosities in the source apertures were negative. This means that the background level is actually higher than that of the source. Inspection of the images shows that many contain point sources distributed randomly outside the source aperture. This could be consistent with ejections of XRBs from star clusters in M51. From the diameter of our image ($\sim$ 425 pc) it is possible that the point sources we see have been ejected from clusters up to $\sim$ 200 pc, similar to results from other groups. This could contribute to the lack of both bright and faint XRBs within 1.5\arcsec\ ($\sim$ 60 pc) of clusters.

\begin{figure}[!ht]
\plotone{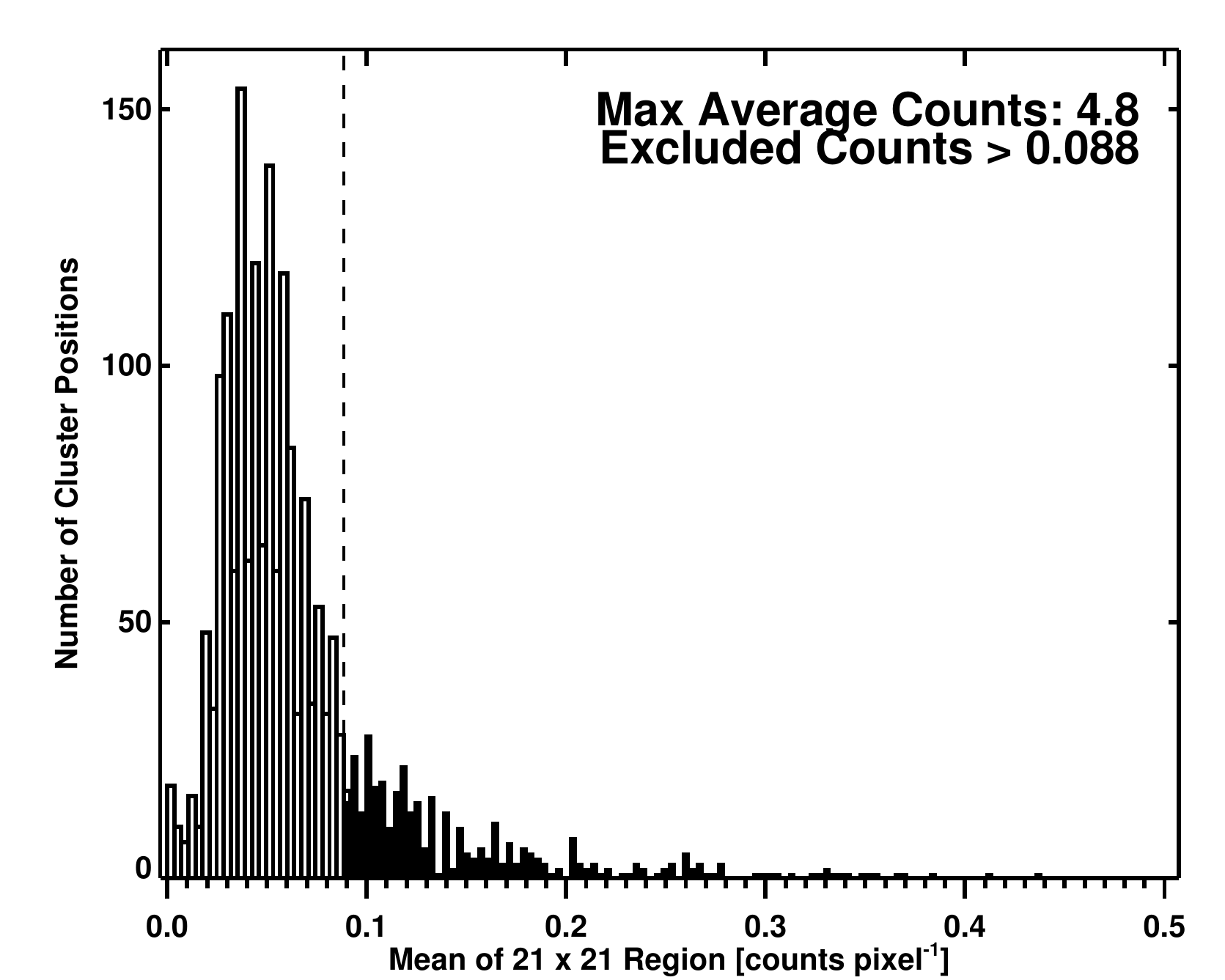}
\caption{Histogram representing the mean value of a pixel in the entire $21 \times 21$ region for each young star cluster stacked. Some bins overlap due to the large bin size we have chosen for illustration purposes. The long tail constitutes star clusters with a large average background value. By removing these from the final stacked image we exclude the bright point sources in the background region.}
\label{fig:histmean}
\end{figure}

In order to be certain that our results were not biased by the numerous bright point sources that are found in the background region, we removed the image(s) from the final stack where these point sources were found. This was accomplished by producing a histogram that represented the mean number of counts per pixel in the entire $21 \times 21$ region for each star cluster stacked. In computing the mean value of each stacked star cluster image, the pixels in the source aperture were all set to zero. This allowed us to retrieve the value of the background without excluding any star clusters that had significant source aperture counts. This may be somewhat skewed since a stacked image with a large mean value might also have a strong source signal. A sample histogram is shown in Figure \ref{fig:histmean} for young clusters. The tail in the histogram has been removed (black portion) since it constitutes the star cluster positions stacked that have large background values and likely contain point sources.

After evaluating every stacked image using this method, we compared the results to those in Figure \ref{fig:stack}. These new images looked uniform and resembled background noise. In contrast to the clusters from Figure \ref{fig:stack}, this approach has resulted in positive source luminosities for most clusters in the full and soft bands. 
However, the opposite is true for old clusters stacked using this new method, likely due to the effect mentioned above where more source signal was removed compared to the background. Nevertheless, this procedure did not result in detections within any of the source apertures.

The fundamental problem with stacking outlined earlier is that diffuse emission in spiral galaxies makes analysis complicated. This X-ray emission comes mainly from HII regions and is a result of supernovae and OB and/or Wolf-Rayet stars heating gas in the interstellar medium to millions of degrees through shocks. In an attempt to reduce the diffuse emission that may be contributing to the background level, we have identified and removed the 150 brightest (H$\alpha$ luminosity $>10^{38}$ \es) HII regions taken from the catalogue of \citet{lee07-11}. We do this by excluding the star clusters associated with these identified HII regions from the final stack. Again, it is possible that by excluding a star cluster this way we may be eliminating signal from our source aperture in the process. We have also applied the \HSTt~offsets mentioned in section \ref{sec:opt} to this HII region catalogue since it uses the same data.

All stacked images produced using this method were again consistent with non-detections. Comparing to Figure \ref{fig:stack}, we found that numerous point sources and diffuse areas of X-ray emission were removed. This method, however, does not eliminate all point sources from the background as did the previous approach. For both all and young cluster types our source luminosities are now positive, which would be expected since young clusters are generally located near HII regions. 
Conversely, we find no change to intermediate-age clusters. The old cluster sample soft and full-band source luminosities changed from positive to negative.
Lastly, we applied both methods mentioned above to the original stacked images to see what effect, if any, this would have on producing a source signal. No significant changes were observed as we retrieved upper limits for all stacked images.

All the methods used above in stacking star clusters overlook the fact that only a small percentage of clusters may have faint XRBs. This would systematically average out any signal that could be detected as star clusters with no XRBs are stacked. In an effort to address this issue, we have used the findings mentioned in section \ref{sec:intro} that state bright, compact, and massive clusters are known to be more likely to host XRBs. Using V magnitudes and effective radii of M51 clusters from \citetalias{hwang04-08}, we stacked the brightest and most compact (assumed small R$_{eff}$) $10$\% of clusters. This was done for the complete cluster sample as well as the various age ranges (Table \ref{tab:scgroups}). The results were unchanged as non-detections followed for all attempts. Another issue that arises when stacking star clusters at large off-axis angles is the increased size of the PSF, which would ultimately skew the flux calculations based on fixed aperture size and background region used throughout. Ideally, we would like to choose the aperture size and background region based on the PSF size at the location of each star cluster that was used in a stacked image. However, because we compute luminosities on the final stacked images, we require all the apertures and background regions to be of identical size. We addressed this problem by stacking cluster types in various off-axis angle ranges where the PSF-size is approximately constant ($0\arcmin-2 \arcmin, 2\arcmin-3 \arcmin, 3\arcmin-4 \arcmin$) and adjusting the source and background regions according to PSF size.
The result was an increase to our upper limits by $\sim$ 15\% that is a consequence of the decrease in counts N (fewer clusters stacked), which increases the noise (1 /$\sqrt{N}$) and hence our upper limits.

\begin{figure*}
\epsscale{1}
\plotone{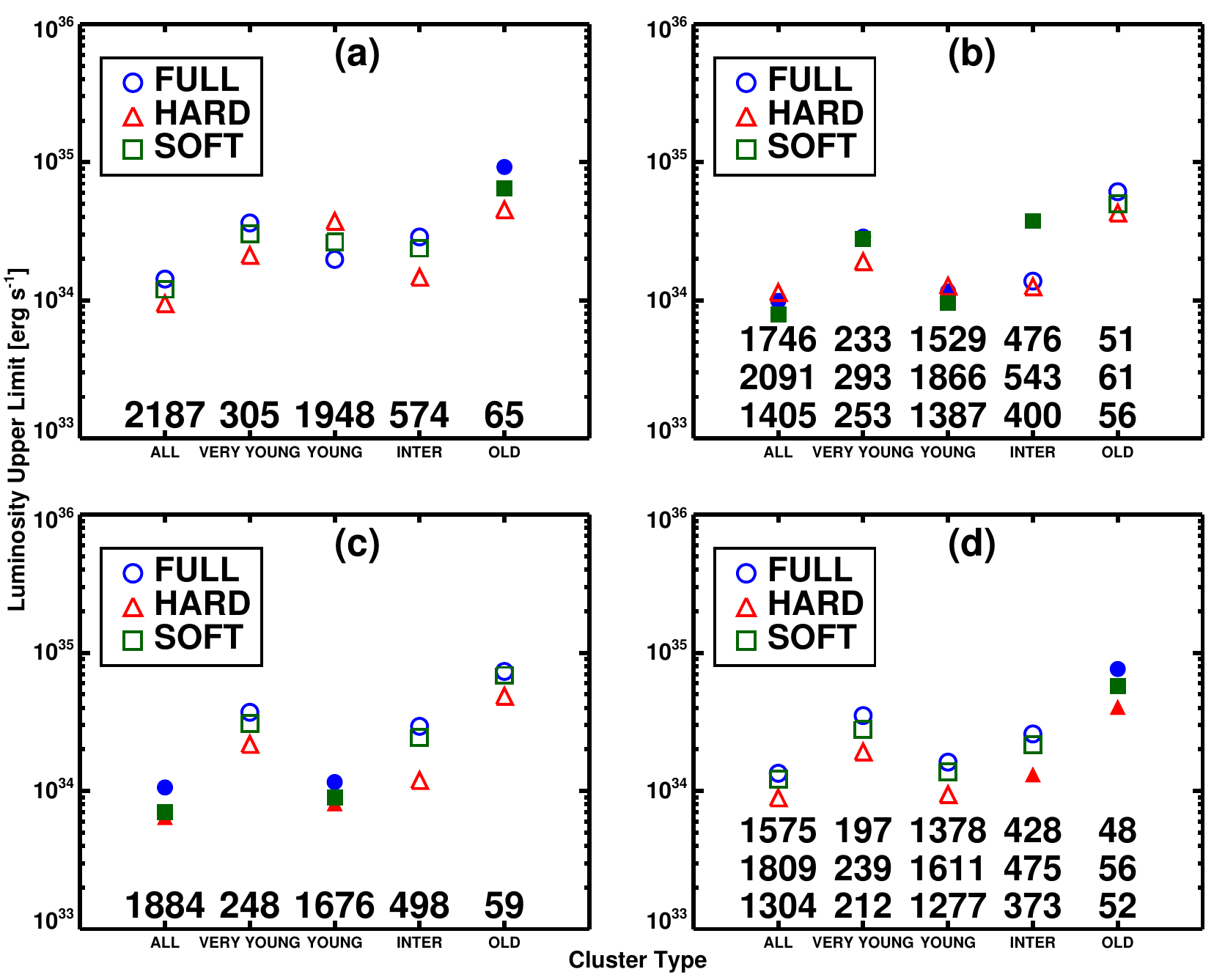}
\caption{Upper limits in \es\ plotted against cluster type in the full (blue circles), hard (red triangles), and soft (green squares) bands. Filled shapes correspond to positive source luminosities whereas hollow shapes represent negative source luminosities. The values above the x-axes in panels (a) and (c) are the number of clusters of each type stacked whereas (b) and (d) show the full, hard, and soft-band numbers from top to bottom respectively. Panel (a) shows the results with no modifications (Figure \ref{fig:stack}), and (b) shows the results after removing the tail of the histogram (see Figure \ref{fig:histmean}) from the stacked image, which represented the star clusters with the largest average pixel values. Panel (c) corresponds to the removal of the 150 brightest (H$\alpha$ luminosity $>10^{38}$ \es) HII regions from the stack. Panel (d) shows the outcome of applying both methods. The smaller number of clusters present in a stacked image led to larger upper limits since a decrease in counts N increases the noise (1 /$\sqrt{N}$).}
\label{fig:clim}
\end{figure*}

Figure \ref{fig:clim} shows the upper limits in \es\ plotted against the type of cluster that was stacked in the full (blue circles), hard (red triangles), and soft (green squares) bands.
The results from all four methods mentioned above (stacking with no modifications as in Figure \ref{fig:stack}, stacking using the method outlined in Figure \ref{fig:histmean}, stacking after removing the 150 brightest (H$\alpha$ luminosity $>10^{38}$ \es) HII regions from the stack, and stacking applying both of the last two methods) are presented in Table \ref{tab:stackvals}. For all the luminosities calculated, a foreground absorption term (assuming a Galactic n$_{H}$ value of $1.52 \times 10^{20}$ cm$^{-2}$ and power-law spectrum of $\Gamma = 1.7$) would increase our upper limits by $\sim$ 4\%.

\subsubsection{Upper Limits}

An interesting result from Figure \ref{fig:clim} is that the young cluster soft and hard-band limits in panel (a) are above the full-band value. Intuitively, the sum of the soft and hard-band counts should be equal to the counts in the full-band, but this is not true for the luminosity (flux). This discrepancy is a consequence of the procedure used to obtain the fluxed images in each energy band. From Section \ref{sec:xraydata}, we mentioned that a monoenergy was used to calculate the exposure map in order to obtain a fluxed image in each energy band. This is not as accurate as using a weighted spectrum file for the different energy bands because the effective area is not constant throughout the entire energy range. Due to the strong energy dependence of effective area, systematic uncertainties are introduced that alter the flux value in each pixel from its true value.\footnote{\url{http://cxc.harvard.edu/ciao/threads/spectral\textunderscore weights/index.html\#intro}}
Since the monoenergy used for the hard-band was $\sim$ 4.7 keV, all energies up to the 8 keV limit in this band would have had their flux underestimated and consequently introduced a smaller background value. A similar argument can be made for the monoenergy of $\sim$ 0.93 keV used for the soft-band.
These smaller background values led to a larger background-subtracted source flux and consequently luminosity. The uncertainties in luminosity were determined using the conversion factor mentioned in Section \ref{sec:results}, which is the ratio of source luminosity to source counts. Since this factor increases with source luminosity it results in much larger upper limits.
This occurs only for the young clusters because we expect there to be more diffuse emission concentrated near them. Nevertheless, the variations are small and do not pose any significant problems with the analysis.

The upper limits vary based on numerous characteristics of the energy band, cluster type, and stacking method. In panel (c) of Figure \ref{fig:clim}, the soft-band upper limit for young clusters increases above that of the hard-band. This change results from the fact we have removed HII regions from the images, which are known to emit soft X-rays around $kT_{e} = 0.6 \pm 0.2$ keV \citep{tullmann12-09}.
This left mostly harder emission features in the soft-band. From our explanation above, this should underestimate the background even more and lead to an overestimation of the upper limit.

\section{Discussion}

\subsection{Upper Limits}

\citet{liu04-06} state that most LMXBs spend their time in quiescence at luminosities of $<$ 10$^{34}$ \es, meaning that our upper limits are just above that threshold to detect this faint population. Our lowest upper limit was found for all clusters in the hard band, $6.5 \times 10^{33}$ \es\ per cluster. However, this is only a 1.3$\sigma$ upper limit and occurs in the hard band, where we don't necessarily expect faint XRBs to be found.

The general trend in Figure \ref{fig:clim} follows that the upper limit and number of clusters stacked are inversely proportional.
All our upper limits are below the threshold of $\approx 5 \times 10^{36}$ \es, which was the minimum X-ray luminosity from the catalogue of point sources (\citetalias{liu01-11}).
Our largest upper limit was from panel (a) of Figure \ref{fig:clim} for the old cluster population in the full-band, $\approx 9 \times 10^{34}$ \es, more than an order of magnitude below the threshold luminosity of catalogued point sources. This means we have the highest probability of observing bright sources (LMXBs) in old clusters compared to other cluster types.
Future deeper (750 ks) observations of M51 with \Chandra\ will be able to detect faint XRBs in star clusters.

\subsection{Comparisons to the Milky Way}

\subsubsection{Globular Clusters}

To put our analysis into context, we need to compare our results to those found for other galaxies, such as the Milky Way (MW). From a catalogue of LMXBs in the MW \citep{liu06-07} we selected the 13 LMXBs associated with GCs. Using distances from the GC catalogue by \citet{harris10-96}[2010 edition], we find a total X-ray luminosity of GC-LMXBs in the MW to be $\sim 1.5 \times 10^{38}$ \es. From the 157 GCs in the MW reported by \citet{harris10-96}[2010 edition] we determine the average X-ray luminosity of MW GCs to be $\approx 10^{36}$ \es.

From Figures \ref{fig:m51optical} and \ref{fig:sccc} we assumed that matches 2, 12, 18, and 19 are all heavily reddened. In addition, match 20 was assumed to be an absorbed background source (AGN) because it is located far from the extent of the spiral arms and has red $V-I$ colors. Therefore only match 4 is likely to be a GC-LMXB (even though it is classified as an absorbed source). 
This gives us a total X-ray luminosity of $\sim 3.15 \times 10^{37}$ \es, meaning that the 65 GCs in M51 currently have an average X-ray luminosity of $\sim 5 \times 10^{35}$ \es.
This is half the value obtained for the average luminosity of MW GCs while the total X-ray luminosity is almost an order of magnitude smaller. This suggests we have not yet observed a fraction of LMXBs in M51. 
From this approximation we can say that either a few (quiescent) bright ($\approx 10^{35-36}$ \es) or a large population of faint ($< 10^{35}$ \es) GC-LMXBs should exist in M51. The more likely scenario would be that a combination of both (quiescent) bright and faint LMXBs are present in M51's GCs to account for the lower average X-ray luminosity compared to the MW. It is also possible that the GC catalogue in M51 is incomplete.

\subsubsection{Young Clusters}

The connection between HMXBs and SFR has been well-studied in spiral galaxies \citep{grimm03-03, grimm12-03}.
These studies showed that X-ray luminosity increases with SFR, meaning that the higher SFR of M51 (1.8 M$_{\sun}$ yr$^{-1}$; \citealt{colbert02-04}) would suggest that it contains more XRBs (field and cluster) than the MW (0.68 $-$ 1.45 M$_{\sun}$ yr$^{-1}$; \citealt{robitaille02-10}).
Using the relation from \citet{mineo11-11}, we find predicted values of L$_{XRB}$ (M51) = $4.7 \times 10^{39}$ \es\ and L$_{XRB}$ (MW) = $(1.8-3.8) \times 10^{39}$ \es.

However, we want to know the XRBs that are present only in clusters. By matching the YCs in the MW with a catalogue of HMXBs (within 60 pc), we found the total X-ray luminosity of $20$ YC-HMXBs to be $\sim 2.7 \times 10^{37}$ \es. For the MW, the YC population is not complete past a few kiloparsecs (kpc) due to interstellar extinction. \citet{dias07-02}[2012 edition] have confirmed $\approx$ 2100 YCs in the MW, meaning that the average X-ray luminosity we would expect from young clusters would be $\sim 1.3 \times 10^{34}$ \es. This value is of course a crude approximation since we observe X-ray point sources further away ($\sim$ 8 kpc; \citealt{grimm-03}) than we do YCs. Of approximately 10 (33\%) HMXBs matched to YCs neither the HMXB or cluster had distance information and therefore were excluded.
From Figures \ref{fig:m51optical} and \ref{fig:sccc} we found numerous X-ray point sources that could be classified as HMXBs. They have a collective luminosity of $\approx 2 \times 10^{39}$ \es, which gives an average X-ray luminosity of $\approx 10^{36}$ \es\ for all young clusters in M51.
This value is two orders of magnitude larger than that calculated for the MW and likely a result of M51's higher SFR. The XRB luminosities calculated using the relation from \citet{mineo11-11} are much larger than those we have calculated since they also include field XRBs along with those located in clusters.
Interestingly, the total X-ray luminosities for young and old clusters in M51 is approximately identical, $\sim 3.15 \times 10^{37}$ \es\ for GC-LMXBs and $\sim 2.7 \times 10^{37}$ \es\ for YC-HMXBs. This is in spite of the GCs being metal-poor and the YCs metal-rich.

These calculations suffer from uncertainty due to the transient nature of XRBs, meaning that it is possible observations were made when some of the brightest XRBs were in a quiescent state. This would skew the number of bright XRBs and ultimately produce lower average X-ray luminosities. 
A partial solution to this problem would be to stack multiple observations (e.g., as \citealt{sivakoff06-08} did for NGC 4697) of M51 in the hope that any transient behaviour would be averaged out.
Our predictions above for GCs in M51 indicate we should observe both more (quiescent) bright and faint LMXBs coincident with star clusters. This will allow us to perform a statistically more meaningful analysis. In addition, the incompleteness of the YC catalogue in the MW allows for a possibly undetected YC-HMXB population.

\subsection{Ejections}

In Section \ref{sec:intro} we summarized the methods of ejecting a binary from a star cluster. These included natal kicks imparted to a NS/BH or from binary-binary or three-body interactions. From section \ref{sec:opt-match} we found that only $\sim 1$\% of M51's star clusters have X-ray emission. Besides detecting faint or transient X-ray sources coincident with clusters, there are other processes that could account for this small fraction.

The most favourable among these would be ejections from star clusters. Models and observations \citep{rangelov11-11, kaaret02-04, kalogera08-06,sepinsky03-05} showed that binaries can be located $\gtrsim$ 200 pc from their parent cluster. Upon inspection of our stacked images in Figure \ref{fig:stack}, we can see that there are an abundance of point sources spread throughout, especially in the stacks of YCs. Our source aperture of radius 1.5$\arcsec$ covers a range of $\sim$ 120 pc from a cluster core. Knowing that most star clusters have cores of only a few pc in size, with GCs being slightly larger, an X-ray point source located within the cluster should appear directly in the centre of an image. It is evident that most of the point sources appear outside the background apertures at distances $>$ 200 pc from the cluster cores. We cannot rule out the possibility that these point sources are XRBs that were in the past located in a star cluster. M51's YCs may have experienced ejections as compact objects were created, however, we have no evidence for this.

\section{Summary \& Conclusions}

We have presented an X-ray analysis of the star clusters in M51 to probe the XRB population. We matched 25 star clusters to 20 X-ray point sources (within 1.5\arcsec) and found that the majority were representative of HMXBs while those remaining were characteristic of SNRs. Of the clusters in M51, $\sim$ 1\% have detected X-ray point sources. \citet{sivakoff05-07} determined $\sim$ 4\% of GCs in 11 Virgo Cluster early-type galaxies had X-ray emission, which was larger than the $\sim$ 0.5\% we found when using the same matching radius of only 1\arcsec. The discrepancy could have been a result of the short lifetimes of HMXBs combined with the larger time period available for LMXB formation in GCs. Our stacking of star cluster positions in an X-ray image was completed to investigate the low-luminosity end of the XLF. Stacking various cluster types in the full, hard, and soft energy bands resulted in non-detections throughout. The largest upper limit (1.3$\sigma$) was found to be $9.23 \times 10^{34}$ \es\ for the average GC in the full-band while the lowest upper limit was $6.46 \times 10^{33}$ \es\ for all clusters in the hard-band. The latter limit indicates that any source detections would have to consist of faint XRBs. The former limit would require that GCs have the largest luminosities in comparison with other cluster types. However, this was biased by the small sample size of GCs. 

We implemented a number of methods in order to reduce the background level to identify any source signal that was present. This included removing bright HII regions and all point sources that were present in the background region. In addition, we stacked clusters that were brightest and most compact since they are more likely to host XRBs. This prevented any possible signal from being averaged out by stacking clusters with no X-ray emission. These methods all resulted in non-detections.

The total X-ray luminosities of young and old clusters in M51 are $\approx 2 \times 10^{39}$ \es\ and $\approx 10^{37}$ \es\ respectively. Comparing this to clusters in the MW, there should be a population of undetected transient and faint XRBs in M51's GC population. In contrast, M51's YC population has more X-ray emission than that of the MW's due to its higher SFR. Average X-ray luminosities of young and old clusters in M51 were $\approx 10^{36}$ \es\ and $\approx 5 \times 10^{35}$ \es\ respectively.
From inspection of the stacked images it follows that numerous bright point sources could have been ejected from clusters in the past. If the same is true for faint XRBs then they would represent the low-luminosity sources we are searching for. Ultimately, we conclude that the data is not deep enough to identify faint XRBs. With upcoming $750$ ks \Chandra\ observations of M51 in 2012 September (P.I.: K. D. Kuntz), we aim to classify the low-luminosity XRBs present in star clusters. Comparisons with other spiral galaxies and a contrast with ellipticals is necessary to develop a complete understanding of faint extragalactic X-ray sources.

\acknowledgements
We thank the referee for helpful comments that improved the paper. We also thank G. R. Sivakoff, R. Chandar, and C. O. Heinke for useful discussions and A. E. Hornschemeier for comments that improved the manuscript. Support for this work was provided by Discovery Grants from the Natural Sciences and Engineering Research Council of Canada and by Ontario Early Researcher Awards.
We acknowledge the following archives: the Hubble Legacy Archive (\url{hla.stsci.edu}), Chandra Data Archive (\url{cda.harvard.edu/chaser}), and 2MASS (\url{ipac.caltech.edu/2mass}). \\
\indent \emph{Facilities:} HST (ACS), CXO (ACIS), 2MASS

\bibliographystyle{aa}

\clearpage

\begin{landscape}
\begin{deluxetable*}{ c  c  c  c  c  c  c  c  c  c  c  c  c }
\tabletypesize{\scriptsize}
\tablecaption{Star Cluster \& X-ray Point Source Matches \label{tab:scxrmatches}}
\tablewidth{0pt}
\tablehead{\multicolumn{3}{c|}{Star Cluster Positions} & \multicolumn{6}{c}{X-ray Data\tablenotemark{a}} \\
\cline{1-3} \cline{4-9} \\
\colhead{ID Number\tablenotemark{b}} & \colhead{RA (J2000)} & \colhead{Dec (J2000)} & \colhead{RA (J2000)} & \colhead{Dec (J2000)} & \colhead{PErr\tablenotemark{c} ($\arcsec$)} & \colhead{$\sigma$\tablenotemark{d}} & \colhead{OAA\tablenotemark{e} ($\arcmin$)} & \colhead{L$_{X}$\tablenotemark{f}} & \colhead{Group ID\tablenotemark{g}} & \colhead{Group Size} & \colhead{Separation ($\arcsec$)} & \colhead{Fig. \ref{fig:m51optical} ID } }

\startdata
    34381 & 202.434767 & 47.1723307 & 202.4343917 & 47.1724798 & 1.1 & 2.9 & 2.2 & 0.13 & 1 & 2 & 1.1 & 1
 \\ 34740 & 202.434706 & 47.1726168 & 202.4343917 & 47.1724798 & 1.1 & 2.9 & 2.2 & 0.13 & 1 & 2 & 0.9 & 1b
 \\ 54644 & 202.4765761 & 47.1890009 & 202.4764709 & 47.1893187 & 1.1 & 5.3 & 0.9 & 0.31 & 2 & 4 & 1.2 & 2
 \\ 54838 & 202.4765456 & 47.1892565 & 202.4764709 & 47.1893187 & 1.1 & 5.3 & 0.9 & 0.31 & 2 & 4 & 0.3 & 2b
 \\ 55245 & 202.4762099 & 47.1896418 & 202.4764709 & 47.1893187 & 1.1 & 5.3 & 0.9 & 0.31 & 2 & 4 & 1.3 & 2c
 \\ 55656 & 202.4763014 & 47.1892412 & 202.4764709 & 47.1893187 & 1.1 & 5.3 & 0.9 & 0.31 & 2 & 4 & 0.5 & 2d
 \\ 94114 & 202.5175154 & 47.2222231 & 202.5179125 & 47.2224409 & 1.1 & 102.7 & 3 & 8 & 3 & 2 & 1.2 & 3
 \\ 94739 & 202.5177595 & 47.2223375 & 202.5179125 & 47.2224409 & 1.1 & 102.7 & 3 & 8 & 3 & 2 & 0.5 & 3b
 \\ 3791 & 202.4082472 & 47.1423166 & 202.4080917 & 47.1422104 & 1.2 & 3.8 & 3.9 & 0.31 & \nodata & \nodata & 0.5 & 4
 \\ 14773 & 202.4772017 & 47.1555879 & 202.4772 & 47.1559992 & 1.1 & 25 & 1.4 & 1.22 & \nodata & \nodata & 1.5 & 5
 \\ 15540 & 202.4752028 & 47.1562174 & 202.4749209 & 47.1565409 & 1.1 & 4.9 & 1.4 & 0.2 & \nodata & \nodata & 1.4 & 6
 \\ 19046 & 202.395796 & 47.1598566 & 202.3953917 & 47.1596604 & 1.2 & 7.6 & 4 & 0.35 & \nodata & \nodata & 1.2 & 7
 \\ 37133 & 202.4754011 & 47.1754358 & 202.4748209 & 47.1754992 & 1.1 & 6.3 & 0.6 & 0.29 & \nodata & \nodata & 1.4 & 8
 \\ 38010 & 202.5246565 & 47.1758364 & 202.52425 & 47.1755798 & 1.1 & 4.7 & 1.5 & 0.2 & \nodata & \nodata & 1.4 & 9
 \\ 59647 & 202.4299605 & 47.1930368 & 202.4303834 & 47.1929798 & 1.1 & 181.4 & 2.5 & 15.03 & \nodata & \nodata & 1.1 & 10
 \\ 61929 & 202.4760725 & 47.1938722 & 202.4757 & 47.1936492 & 1.1 & 19.7 & 1.1 & 1.85 & \nodata & \nodata & 1.2 & 11
 \\ 63390 & 202.4824812 & 47.1958063 & 202.4827334 & 47.1956992 & 1.1 & 7.5 & 1.1 & 0.46 & \nodata & \nodata & 0.7 & 12
 \\ 66404 & 202.433531 & 47.1989267 & 202.4335834 & 47.1989909 & 1.1 & 8.1 & 2.6 & 0.35 & \nodata & \nodata & 0.3 & 13
 \\ 80844 & 202.5089095 & 47.2103746 & 202.50865 & 47.2106492 & 1.1 & 5.5 & 2.2 & 0.26 & \nodata & \nodata & 1.2 & 14
 \\ 88301 & 202.4479658 & 47.2168367 & 202.448 & 47.2169298 & 1.1 & 11.1 & 2.9 & 0.54 & \nodata & \nodata & 0.3 & 15
 \\ 94501 & 202.4932845 & 47.2225893 & 202.4931125 & 47.2223798 & 1.1 & 22.6 & 2.8 & 1.37 & \nodata & \nodata & 0.9 & 16
 \\ 112533 & 202.4305861 & 47.2571962 & 202.43075 & 47.2569409 & 1.1 & 35.8 & 5.4 & 3.48 & \nodata & \nodata & 1.0 & 17
 \\ 113286 & 202.4991438 & 47.2611444 & 202.4991625 & 47.2613104 & 1.2 & 11.7 & 5.1 & 1.49 & \nodata & \nodata & 0.6 & 18
 \\ 115180 & 202.4894392 & 47.2699488 & 202.4895834 & 47.2699992 & 1.1 & 29.2 & 5.6 & 3.96 & \nodata & \nodata & 0.4 & 19
 \\ 115321 & 202.4081099 & 47.2705744 & 202.4082709 & 47.2705798 & 1.9 & 7.8 & 6.6 & 0.72 & \nodata & \nodata & 0.4 & 20
 \\ 
\enddata
\tablecomments{The chance coincidence probability between matched sources is $\sim$ 50\%, where the method is summarized in Section \ref{sec:xrmatches}.}

\tablenotetext{a}{Taken from the catalogue of \citet{liu01-11}.}
\tablenotetext{b}{From the catalogue of \citet{hwang04-08}.}
\tablenotetext{c}{Combination of source positional uncertainty from \citet{liu01-11} and \Chandra\ astrometric error taken to be $\sim$0.5$\arcsec$ for the small count regime \citep{sivakoff05-07}.}
\tablenotetext{d}{Detection significance from $\texttt{wavdetect}$.}
\tablenotetext{e}{Off-axis angle from \citet{liu01-11}.}
\tablenotetext{f}{Units of 10$^{38}$ \es\ in the 0.3 $-$ 8 keV band.}
\tablenotetext{g}{Group ID identifies multiple star clusters (Group Size) that were matched to the same X-ray point source.}
\end{deluxetable*}
\clearpage
\end{landscape}

\clearpage
\begin{landscape}
\begin{deluxetable*}{c c c c c c c c c c c c c}
\tabletypesize{\scriptsize}
\tablecaption{Stacked Image Properties \label{tab:stackvals}}
\tablecolumns{13}
\tablewidth{0pt}
\tablehead{
\colhead{Cluster Type\tablenotemark{a}} &
\multicolumn{4}{c|}{FULL} &
\multicolumn{4}{c|}{HARD} &
\multicolumn{4}{c}{SOFT} \\
\cline{2-5} \cline{6-9} \cline{10-13}
\colhead{} & 
\colhead{Number} &
\colhead{Upper Limit\tablenotemark{b}} &
\colhead{Sky\tablenotemark{c}} &
\colhead{Sky Sigma\tablenotemark{d}} &
\colhead{Number} &
\colhead{Upper Limit\tablenotemark{b}} &
\colhead{Sky\tablenotemark{c}} &
\colhead{Sky Sigma\tablenotemark{d}} &
\colhead{Number} &
\colhead{Upper Limit\tablenotemark{b}} &
\colhead{Sky\tablenotemark{c}} &
\colhead{Sky Sigma\tablenotemark{d}} \\
\cline{3-5} \cline{7-9} \cline{11-13}
\colhead{} & 
\colhead{} &
\multicolumn{3}{c}{(10$^{34}$ \es)} &
\colhead{} &
\multicolumn{3}{c}{(10$^{34}$ \es)} &
\colhead{} &
\multicolumn{3}{c}{(10$^{34}$ \es)}
}
\startdata
All & 2187 & 1.425 & 11.034 & 0.05 & 2187 & 0.945 & 2.71 & 0.029 & 2187 & 1.204 & 8.158 & 0.042
 \\ Young & 1948 & 1.979 & 10.459 & 0.052 & 1948 & 3.734 & 2.766 & 0.032 & 1948 & 2.643 & 7.508 & 0.044
 \\ Very Young & 305 & 3.633 & 9.169 & 0.129 & 305 & 2.113 & 2.336 & 0.072 & 305 & 3.044 & 6.895 & 0.108
 \\ Inter & 574 & 2.876 & 10.941 & 0.098 & 574 & 1.468 & 2.461 & 0.053 & 574 & 2.385 & 8.426 & 0.082
 \\ Old & 65 & 9.234 & 9.948 & 0.296 & 65 & 4.519 & 2.392 & 0.155 & 65 & 6.492 & 7.903 & 0.255
  \\ \cutinhead{Method \#2\tablenotemark{f}}
 \\ All & 1746 & 0.992 & 5.226 & 0.038 & 2091 & 1.138 & 2.061 & 0.025 & 1405 & 0.794 & 2.799 & 0.03
 \\ Young & 1529 & 1.161 & 5.208 & 0.04 & 1866 & 1.265 & 2.067 & 0.026 & 1387 & 0.955 & 3.122 & 0.032
 \\ Very Young & 233 & 2.897 & 5.706 & 0.109 & 293 & 1.902 & 2.086 & 0.067 & 253 & 2.779 & 4.31 & 0.087
 \\ Inter & 476 & 1.378 & 5.236 & 0.072 & 543 & 1.25 & 2.059 & 0.049 & 400 & 3.752 & 2.898 & 0.057
 \\ Old & 51 & 6.115 & 4.584 & 0.212 & 61 & 4.279 & 1.982 & 0.147 & 56 & 4.981 & 3.411 & 0.168
   \\ \cutinhead{Method \#3\tablenotemark{g}}
 \\ All & 1884 & 1.059 & 8.939 & 0.05 & 1884 & 0.646 & 2.533 & 0.031 & 1884 & 0.703 & 6.472 & 0.041
 \\ Young & 1676 & 1.159 & 8.69 & 0.052 & 1676 & 0.813 & 2.531 & 0.032 & 1676 & 0.893 & 6.207 & 0.043
 \\ Very Young & 248 & 3.709 & 8.574 & 0.135 & 248 & 2.172 & 2.326 & 0.08 & 248 & 3.069 & 6.335 & 0.112
 \\ Inter & 498 & 2.935 & 9.558 & 0.097 & 498 & 1.189 & 2.345 & 0.055 & 498 & 2.45 & 7.337 & 0.081
 \\ Old & 59 & 7.33 & 7.277 & 0.245 & 59 & 4.845 & 2.153 & 0.156 & 59 & 6.859 & 5.143 & 0.2
   \\ \cutinhead{Method \#4\tablenotemark{h}}
 \\ All & 1575 & 1.342 & 7.306 & 0.048 & 1809 & 0.891 & 2.038 & 0.027 & 1304 & 1.212 & 5.245 & 0.043
 \\ Young & 1378 & 1.616 & 7.384 & 0.053 & 1611 & 0.939 & 2.074 & 0.029 & 1277 & 1.373 & 5.383 & 0.045
 \\ Very Young & 197 & 3.51 & 6.982 & 0.131 & 239 & 1.915 & 1.806 & 0.07 & 212 & 2.78 & 5.124 & 0.104
 \\ Inter & 428 & 2.585 & 7.28 & 0.091 & 475 & 1.318 & 1.937 & 0.051 & 373 & 2.162 & 5.371 & 0.08
 \\ Old & 48 & 7.628 & 6.729 & 0.263 & 56 & 4.062 & 1.758 & 0.144 & 52 & 5.723 & 4.542 & 0.201
 \\ 
\enddata
\tablecomments{A foreground absorption term (assuming a Galactic n$_{H}$ value of $1.52 \times 10^{20}$ cm$^{-2}$ and power-law spectrum of $\Gamma = 1.7$) increases our upper limits by $\sim$ 4\%.}

\tablenotetext{a}{As defined in Table \ref{tab:scgroups}.}
\tablenotetext{b}{On source aperture (radius of 1.5$\arcsec$) luminosity.}
\tablenotetext{c}{The background aperture (outside a radius of 2.5$\arcsec$) luminosity per pixel.}
\tablenotetext{d}{Uncertainty in the background per pixel, determined by averaging the upper and lower limits calculated using equations (9) and (14) of \citet{gehrels04-86}.}
\tablenotetext{f}{High-count background point sources have been removed from the stacked images (see Figure \ref{fig:histmean}).}
\tablenotetext{g}{The 150 brightest (H$\alpha$ luminosity $>10^{38}$ \es) HII regions have been removed from all stacked images.}
\tablenotetext{h}{Methods \#2 and \#3 both applied to all stacked images (see Section \ref{sec:stacking}).}
\end{deluxetable*}
\clearpage
\end{landscape}

\end{document}